\documentclass[]{spie}  %>>> use for US letter paper
\usepackage{amsmath,amsfonts,amssymb}
\usepackage{graphicx}
\usepackage[colorlinks=true, allcolors=blue]{hyperref}

\title{The Primordial Inflation Polarization Explorer (PIPER)}

\author[a,b]{Natalie N. Gandilo}
\author[c]{Peter A. R. Ade}
\author[b]{Dominic Benford}
\author[a]{Charles L. Bennett}
\author[d]{David T. Chuss}
\author[e]{Jessie L. Dotson}
\author[a]{Joseph R. Eimer}
\author[b,f]{Dale J. Fixsen}
\author[g]{Mark Halpern}
\author[h]{Gene Hilton}
\author[g]{Gary F. Hinshaw}
\author[i]{Kent Irwin}
\author[b]{Christine Jhabvala}
\author[j]{Mark Kimball}
\author[b]{Alan Kogut}
\author[b,k]{Luke Lowe}
\author[l]{Jeff J. McMahon}
\author[b]{Timothy M. Miller}
\author[b,k]{Paul Mirel}
\author[b]{S. Harvey Moseley}
\author[b,f]{Samuel Pawlyk}
\author[b,m]{Samelys Rodriguez}
\author[b]{Elmer Sharp III}
\author[j]{Peter Shirron}
\author[a,b]{Johannes G. Staguhn}
\author[j]{Dan F. Sullivan}
\author[b]{Eric R. Switzer}
\author[b,k]{Peter Taraschi}
\author[c]{Carole E. Tucker}
\author[b]{Edward J. Wollack}

\affil[a]{Johns Hopkins University, Baltimore, MD, USA}
\affil[b]{Code 665, NASA Goddard Space Flight Center, Greenbelt, MD, USA}
\affil[c]{Cardiff University, Cardiff, Wales, UK}
\affil[d]{Villanova University, Villanova, PA, USA}
\affil[e]{NASA Ames Research Center, Moffett Field, CA, USA}
\affil[f]{University of Maryland, College Park, MD, USA}
\affil[g]{University of British Columbia, Vancouver, BC, Canada}
\affil[h]{National Institute for Standards and Technology, Boulder, CO, USA}
\affil[i]{Stanford University, Stanford, CA, USA}
\affil[j]{Code 552, NASA Goddard Space Flight Center, Greenbelt, MD, USA}
\affil[k]{Wyle STE, Houston, TX, USA}
\affil[l]{University of Michigan, Ann Arbor, MI, USA}
\affil[m]{ADNET Systems, Inc., Bethesda, MD, USA}

\authorinfo{Send correspondence to Natalie N. Gandilo: E-mail: natalie.n.gandilo@nasa.gov, Telephone: 1 301 286 8310}

\begin{document} 
\maketitle

\begin{abstract}
The Primordial Inflation Polarization ExploreR (PIPER) is a balloon-borne telescope designed to measure the polarization of the Cosmic Microwave Background on large angular scales. PIPER will map 85\% of the sky at 200, 270, 350, and\,600 GHz over a series of 8 conventional balloon flights from the northern and southern hemispheres. The first science flight will use two $32\times40$ arrays of backshort-under-grid transition edge sensors, multiplexed in the time domain, and maintained at 100\,mK by a Continuous Adiabatic Demagnetization Refrigerator. Front-end cryogenic Variable-delay Polarization Modulators provide systematic control by rotating linear to circular polarization at 3 Hz. Twin telescopes allow PIPER to measure Stokes $I$, $Q$, $U$, and $V$ simultaneously. The telescope is maintained at 1.5\,K in an LHe bucket dewar. Cold optics and the lack of a warm window permit sensitivity at the sky-background limit. The ultimate science target is a limit on the tensor-to-scalar ratio of $r\sim0.007$, from the reionization bump to $l\sim300$. PIPER's first flight will be from the Northern hemisphere, and overlap with the CLASS survey at lower frequencies. We describe the current status of the PIPER instrument. 
\end{abstract}


\section{INTRODUCTION}
\label{sec:intro}  
   \begin{figure} [ht]
   \begin{center} 
   \includegraphics[height=6cm]{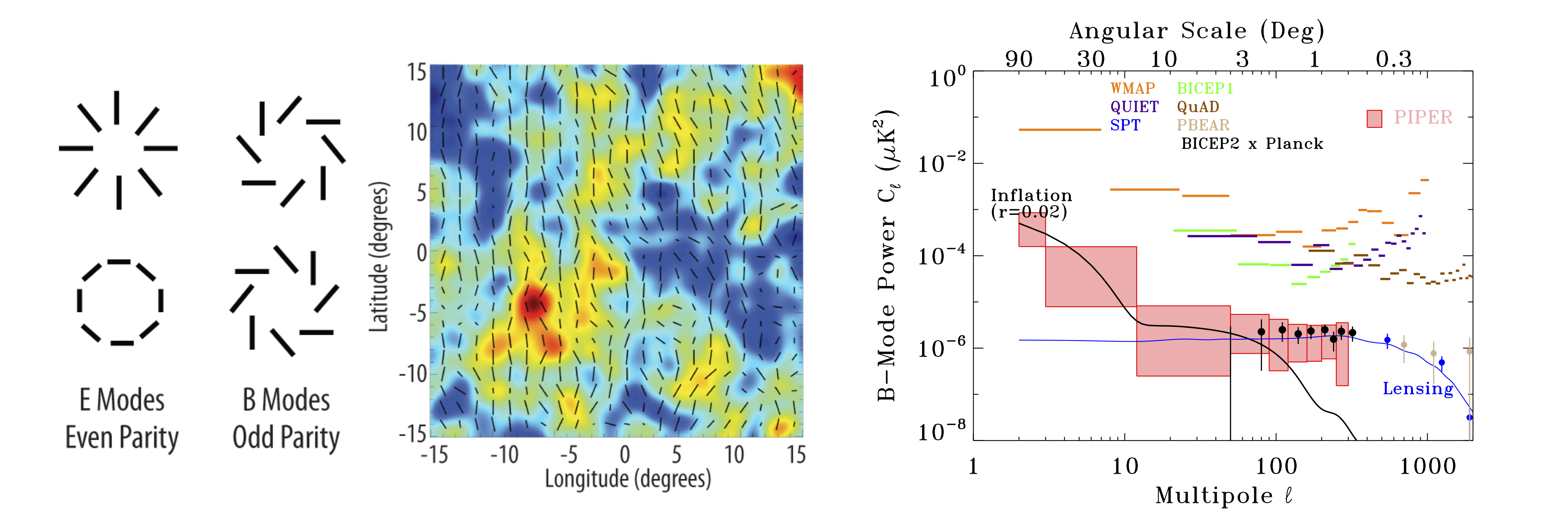}
   \end{center}
   \caption[power spectrum] 
   { \label{fig:piperplaincl} 
Left: The polarization of the CMB can be decomposed into E- and B-modes. Right: B-mode power spectrum assuming a tensor-to-scalar ratio $r=0.02$ from inflationary gravitational waves (black) and gravitational lensing (blue). The quantity being measured, the B-mode power, is plotted against angular scale (large multipole $l$ = small scales). PIPER is the only experiment with the sensitivity to measure both the B-mode signal and the dust foreground at large angular scales, where the inflationary signal is not contaminated by the signal from lensing.}
   \end{figure}
   
Cosmological inflation is the current leading explanation for the observed characteristics of the universe. This theory postulates a period of exponential expansion just after the Big Bang (at $t\approx10^{-35}$ seconds), when the universe expanded by at least 60 e-folds. While the existence of an inflationary epoch can account for observations, no direct evidence yet exists that inflation occurred. The simplest models of inflation predict that the rapid expansion would have produced tensor perturbations in the space-time metric of the universe, i.e. gravitational waves. The current best method for detecting these gravitational waves is to look for the imprint they would have left in the Cosmic Microwave Background (CMB). This imprint comes in the form of a specific pattern, known as B-modes, in the polarization of the CMB.

Figure \ref{fig:piperplaincl} shows the polarization signature that CMB experiments are seeking. The observed pattern of linear polarization can be decomposed into curl-free and curl components, known as E-modes and B-modes, respectively, named in analogy to electric and magnetic fields. While density fluctuations can only produce E-modes, gravitational waves can produce both E-modes and B-modes. Hence a detection of B-modes is considered the ``smoking gun" for inflation. The level of the B-mode signal, parameterized by the ratio, $r$, of tensor-to-scalar perturbations, is directly related to the energy scale of inflation.

The simplest models of inflation predict $r \sim 0.01$, associating it with the physics of Grand Unified Theories at $10^{16}$\,GeV. A B-mode signal would be a direct probe of the universe at energy scales well beyond the reach of high energy particle colliders. At this level, this signal would be detectable by modern CMB experiments. Discovering the signature of inflation in the polarization of the CMB would advance our understanding of fundamental physics and of the origin of our universe.

An inflationary B-mode power spectrum has a characteristic shape, with a ``reionization bump" at large angular scales ($>20^{\circ}$) and a ``recombination peak" at a scale $\sim2^{\circ}$ (Figure \ref{fig:piperplaincl}). The primordial B-mode signal from recombination becomes progressively more contaminated by gravitational lensing of E-modes into B-modes on angular scales smaller than several degrees. The signal is further complicated by the polarized emission of our own galaxy, in particular from interstellar dust and synchrotron radiation. The inflationary B-mode signal is small, at the level of $10^{-9}$\,K, and therefore detecting it requires a high degree of control over systematics and careful removal of galactic foregrounds. 

\begin{figure} [ht]
   \begin{center}
   \includegraphics[height=8cm]{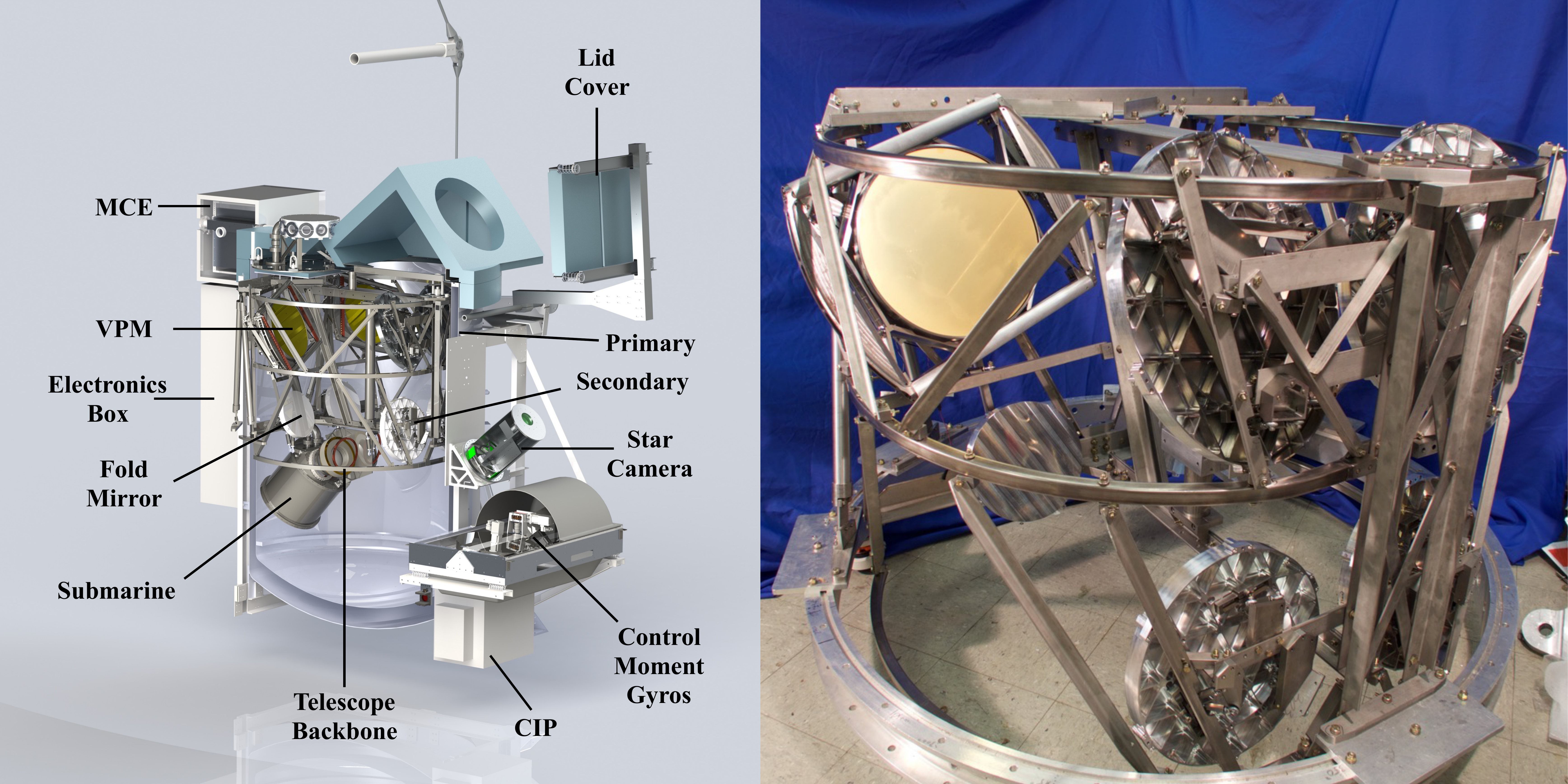}
   \end{center}
   \caption[instrument] 
   { \label{fig:instrument} 
Left: Model of the PIPER payload. Right: PIPER telescope frame. PIPER's twin telescopes are mounted on a fully stainless steel frame in an open-aperture 3500 liter LHe bucket dewar. The telescope is registered to a rigid backbone, and the entire backbone is mounted into the dewar. Only elements of one telescope are labeled; the two telescopes are sagittal mirrors of one another. The first optical element is the VPM, ensuring instrumental polarization cannot be modulated and become systematics. A retractable lid cover protects the detectors on the ground and from the sun. The payload moves in azimuth using a pair of control moment gyros. A star camera provides absolute pointing information.}
   \end{figure} 

Joint analysis \cite{Keck2016} of the data from BICEP2/Keck and \textit{Planck} experiments yields an upper limit of $r<0.07$. However this limit relies heavily on modeling of the foreground emission. Two critical features are needed for any experiment aiming to distinguish an inflationary signal from competing foregrounds. One is the ability to probe large angular scales, where the B-mode signal from the reionization bump does not suffer any contamination from lensing. The other is the ability to measure the dominant dust foreground signal with sensitivity better than \textit{Planck}. PIPER is currently the only instrument capable of doing both, by mapping a large fraction of the sky in four bands, at 200, 270, 350 and 600\,GHz, providing the necessary data to measure the large scale B-modes as well as the dust spectrum.

By flying each frequency of PIPER in both the Southern and Northern Hemisphere, PIPER will be able to map 85\% of the sky. This will allow the shape of the B-mode power spectrum to be measured at angular scales up to $90^{\circ}$, constraining the characteristic reionization bump. Figure \ref{fig:piperplaincl} shows the sensitivity of PIPER compared to an inflationary model with $r=0.02$. PIPER will also measure the polarized dust foreground to a signal-to-noise of better than 10 even for regions of low dust intensity and even for polarization fractions of 10\%. The PIPER design allows a clean measurement of CMB polarization over the full sky, without the need for complicated scan strategies or boresight rotation.

\section{INSTRUMENT}
Figure \ref{fig:instrument} shows the layout of the PIPER instrument. Two twin telescopes are mounted within a 3500 liter open-aperture liquid helium (LHe) bucket dewar used by ARCADE \cite{ARCADE2}. At float altitude the LHe bath temperature is reduced to 1.5\,K, and the telescope is cooled by the evaporating liquid with the aid of superfluid helium pumps. The telescopes are aligned such that one measures Stokes $Q$ and the other simultaneously measures Stokes $U$ in the same position on the sky. A Variable-delay Polarization Modulator (VPM) is the first optical element and modulates the polarization of the light going into each telescope. The sky is imaged onto arrays of TES bolometers housed in a vacuum vessel and cooled to 100\,mK by a Continuous Adiabatic Demagnetization Refrigerator (CADR).

In the fall of 2016, PIPER will have an engineering flight from Fort Sumner, New Mexico. Beginning in the spring of 2017, PIPER will alternately fly from the Northern and Southern hemisphere, with each pair of flights covering 85\% of the sky, for a total of 8 flights. Band-defining filters are placed at the front of the detector package, and can be changed between flights. Anti-reflective (AR) coatings on the lenses and the throw of the VPM mirror are also designed to match the desired passband of the instrument. The angular resolution is approximately $21^{\prime}$ at 200\,GHz, $15^{\prime}$ at 270\,GHz and $14^{\prime}$ at 350 and 600\,GHz.

\begin{figure}
\begin{center}
\includegraphics[height=5.5cm, angle=90]{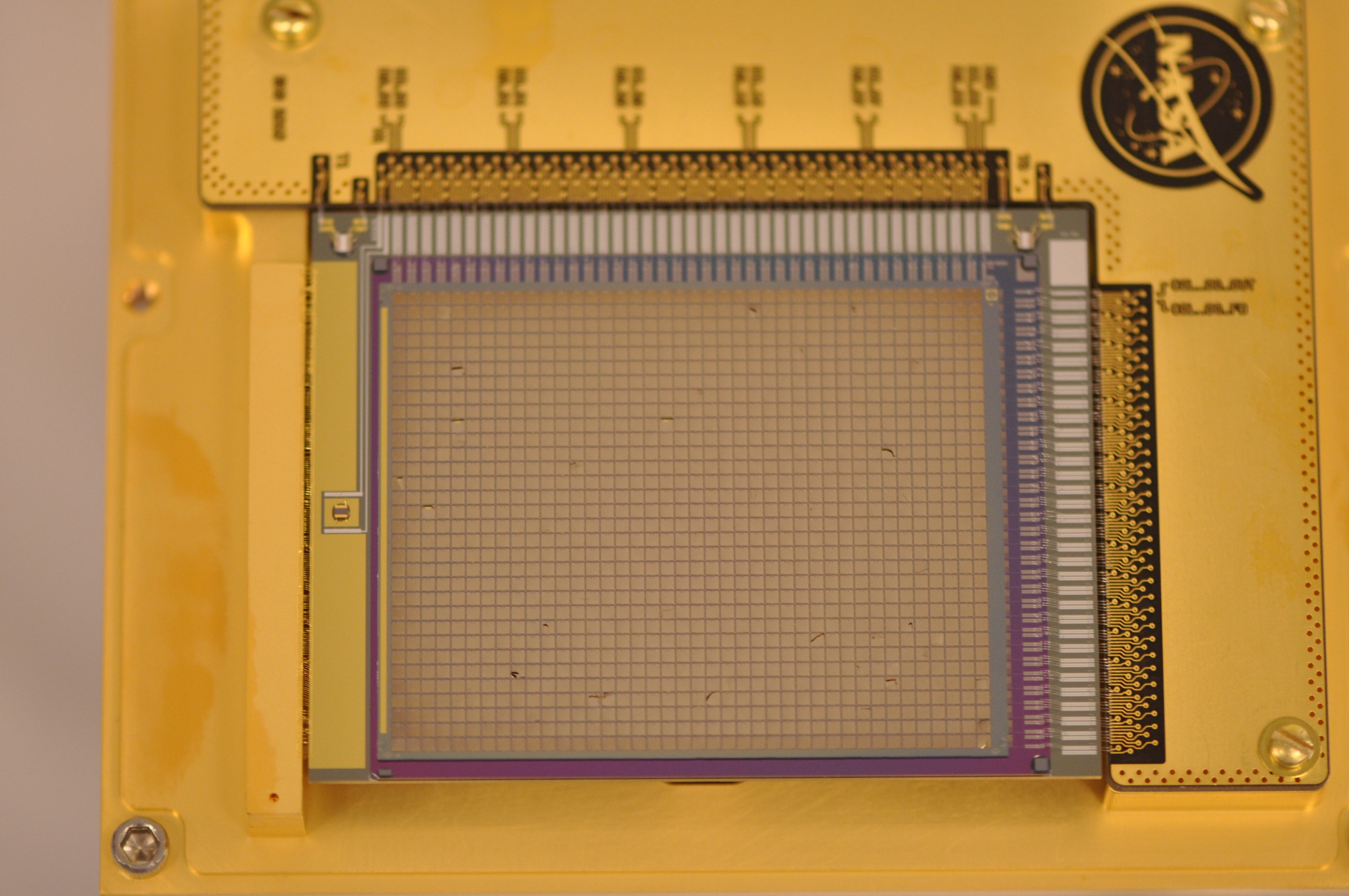}
\includegraphics[height=8cm]{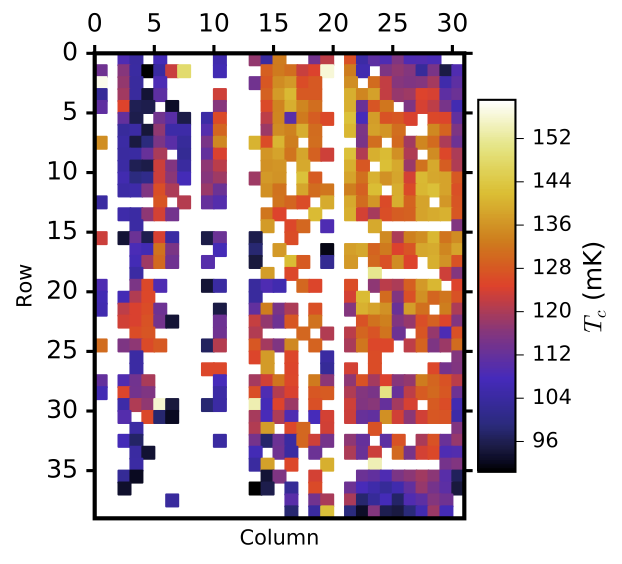}
\end{center}
\caption{Left: $32\times40$ BUG detector array that will be flown on the PIPER engineering flight. Right: Transition temperature, $T_c$, for each pixel across an array.}
\label{fig:detector}
\end{figure} 

\subsection{OPTICS}
PIPER's reflective fore-optics are in an off-axis Gregorian-like configuration with a VPM mirror, primary mirror, folding flat mirror, and secondary mirror. The mirrors are mounted on a stainless steel frame and held at 1.5\,K inside the windowless dewar. Instrumental emission is greatly reduced by having fully cryogenic optics and by the lack of a warm window, allowing PIPER to operate near the CMB photon noise limit at float.

The fore-optics transfer light to the vacuum vessel (``submarine") where re-imaging optics focus the light onto the detectors. A Lyot stop apodizes the illumination of the VPM. Light is slowed by a lens before passing through an analyzer grid with wires oriented at $45^{\circ}$ to the VPM grid. The analyzer grid separates the light into two orthogonal linear polarization components. Each component then passes through a final lens and a band-defining filter before being imaged onto a detector array. The first few flights will have a slightly different design, with optical elements placed outside the submarine (as seen in Figure \ref{fig:instrument}). Following the secondary reflector, the analyzer grid reflects one polarization state to a beam dump, and transmits the other to a cold stop, lens, and filter stack. The light then enters the submarine and is focussed to a band-defining filter and detector array. Lenses are AR-coated with grooved metamaterial developed at the University of Michigan \cite{Datta2013}. The telescope has Strehl ratios above 0.97 over the entire $4.7^{\circ}\times6^{\circ}$ field of view. 

\subsection{DETECTORS}
PIPER's detector arrays consist of $32\times40$ pixels of backshort-under-grid (BUG) transition-edge sensor (TES) bolometers\cite{Jhabvala2014} (see Figure \ref{fig:detector}). A reflective backshort behind each pixel maximizes absorption in the lowest PIPER bands, which contain the CMB signal, while reducing absorption in the highest bands, which contain more of the foreground dust signal. All flights will use the same detector arrays, with absorption efficiency and bandwidths chosen to present constant background loading below saturation. Initial flights will use a single detector array per telescope. In later flights, each of the two telescopes will use two identical detector arrays, each receiving light in a single linear polarization, for a total of 5120 detectors. 
\begin{figure}
\begin{center}
\includegraphics[height=7cm]{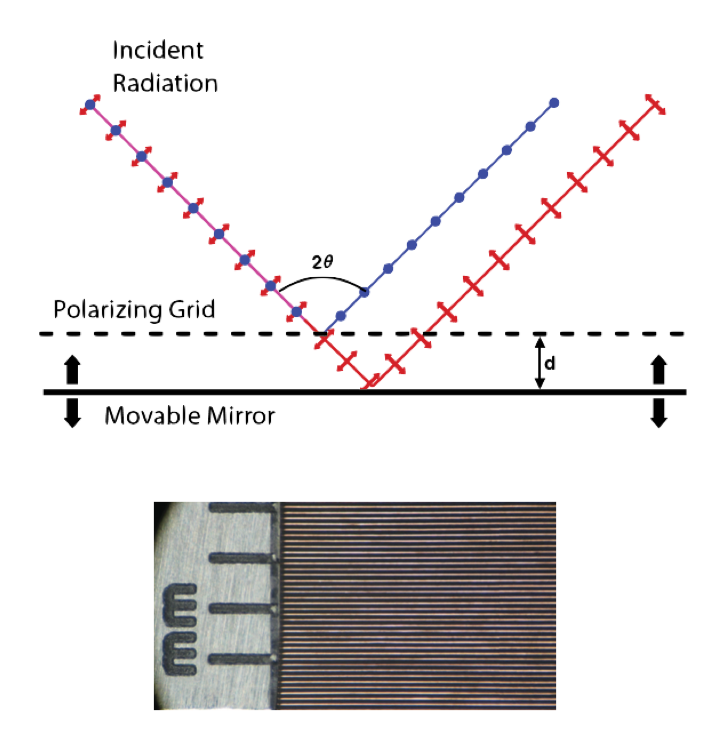}
\includegraphics[height=7cm]{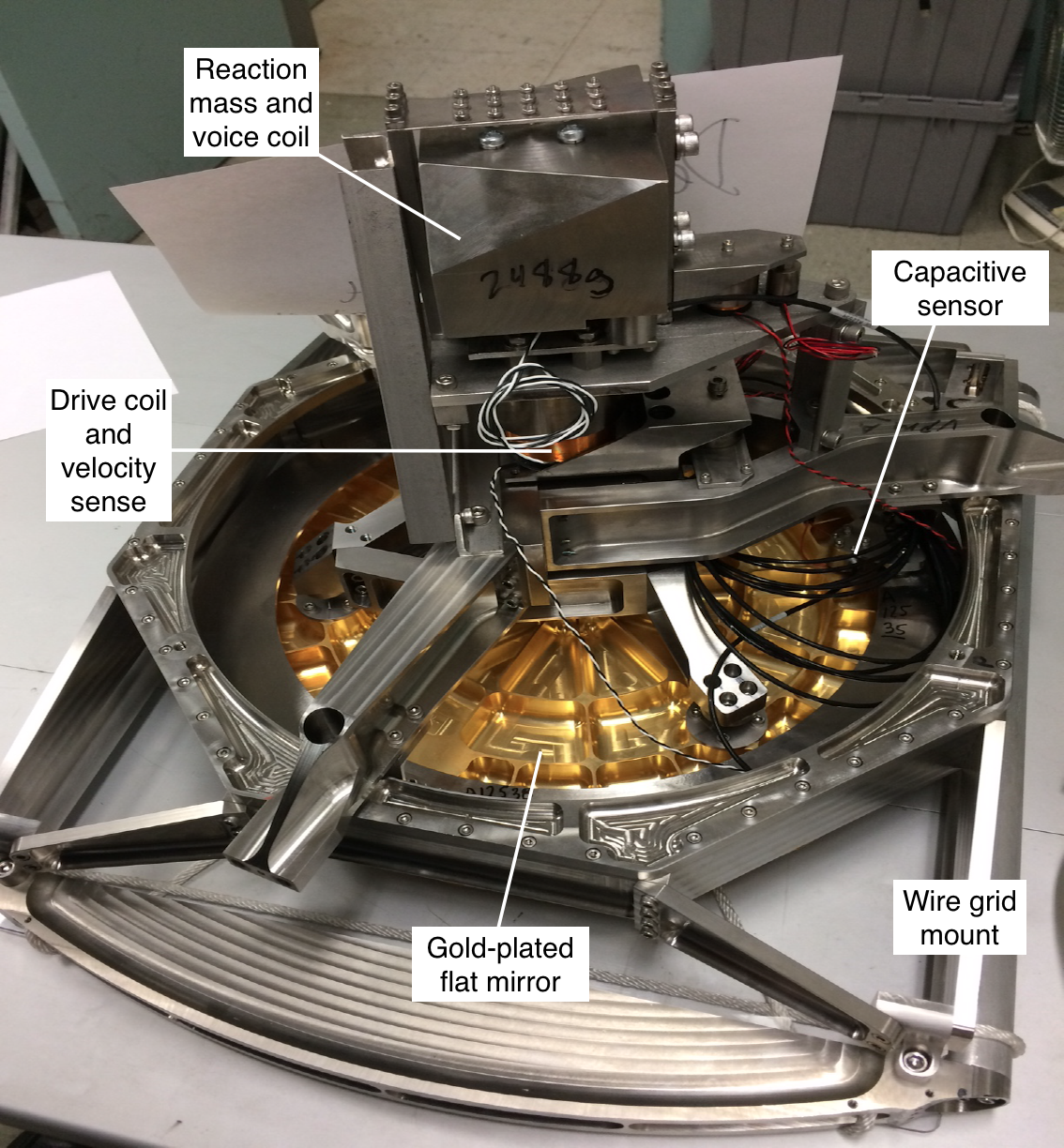}
\end{center}
\caption{Top left: Schematic of VPM operation. A mirror is moved towards and away from a wire grid, introducing a phase delay between light polarized parallel and perpendicular to the grid. Bottom left: Wire grid used in the VPM. Right: Set up of the VPM and voice coil drive, shown without the wire grid installed.}
\label{fig:VPM}
\end{figure}

The detector arrays are indium bump-bonded to time-domain multiplexed superconducting quantum interference devices (SQUIDs) designed by NIST. The TESs and SQUID multiplexer system are read out using Multi-channel Electronics (MCEs) developed by the University of British Columbia \cite{Battistelli2008}.

\subsection{POLARIZATION MODULATION}
\label{sec:vpm}

PIPER makes use of a novel technology for polarization modulation called a Variable-delay Polarization Modulator (VPM) \cite{Chuss2014}. The VPM modulates the polarization of the signal going into each of PIPER's telescopes, cleanly separating polarized from unpolarized emission. As the first optical element, the VPM modulates the sky signal before any instrumental polarization effects can be introduced.

The VPM consists of a grid of 36 $\mu$m wires spaced 115 $\mu$m apart, placed in front of a movable flat mirror. Light polarized parallel to the wires of the grid is reflected from the grid, while light polarized perpendicular to the wires passes through the grid and reflects off the mirror and back through the grid, recombining with the parallel component. A phase delay, $\phi$, is introduced between the parallel and perpendicular components,
\begin{equation}
\phi=\frac{4\pi d}{\lambda}\mbox{cos}\theta,
\end{equation}
where $d$ is the distance between the mirror and the grid, $\lambda$ is the wavelength being observed, and $\theta$ is the incident angle (see Figure \ref{fig:VPM}). This relation assumes that the wavelength is much larger than length scales of the grid geometry.

This phase delay leads to a modulation of local Stokes $U$ and $V$, leaving Stokes $Q$ unchanged. The output of the VPM is split into orthogonal polarization states by an analyzer grid oriented at $45^{\circ}$ to the VPM wire grid. Each detector array measures a single polarization state. PIPER's two telescopes are oriented with their VPM grids at $45^{\circ}$ to each other such that one telescope measures Stokes $Q$ on the sky while the other measures Stokes $U$.

Figure \ref{fig:VPM} shows the VPM assembly. The mirror is moved towards and away from the wire grid by a linear voice coil. The distance between the mirror and grid is measured by capacitive sensors. Another voice coil is used as a velocity sensor, and the motion of the mirror is controlled to have a sinusoidal displacement profile. A duplicate system is attached to a reaction mass, which is actuated in the opposite direction from the mirror in order to minimize the vibrational effect on the telescope. The gold-plated mirror and its mounting structure as well as the voice coil drive assembly are made of stainless steel. The mirror is moved back and forth at 3\,Hz, modulating the sky signal into frequencies well above the $1/f$ knee of the detector noise spectra. 
\begin{figure} [ht]
   \begin{center}
   \includegraphics[height=6cm]{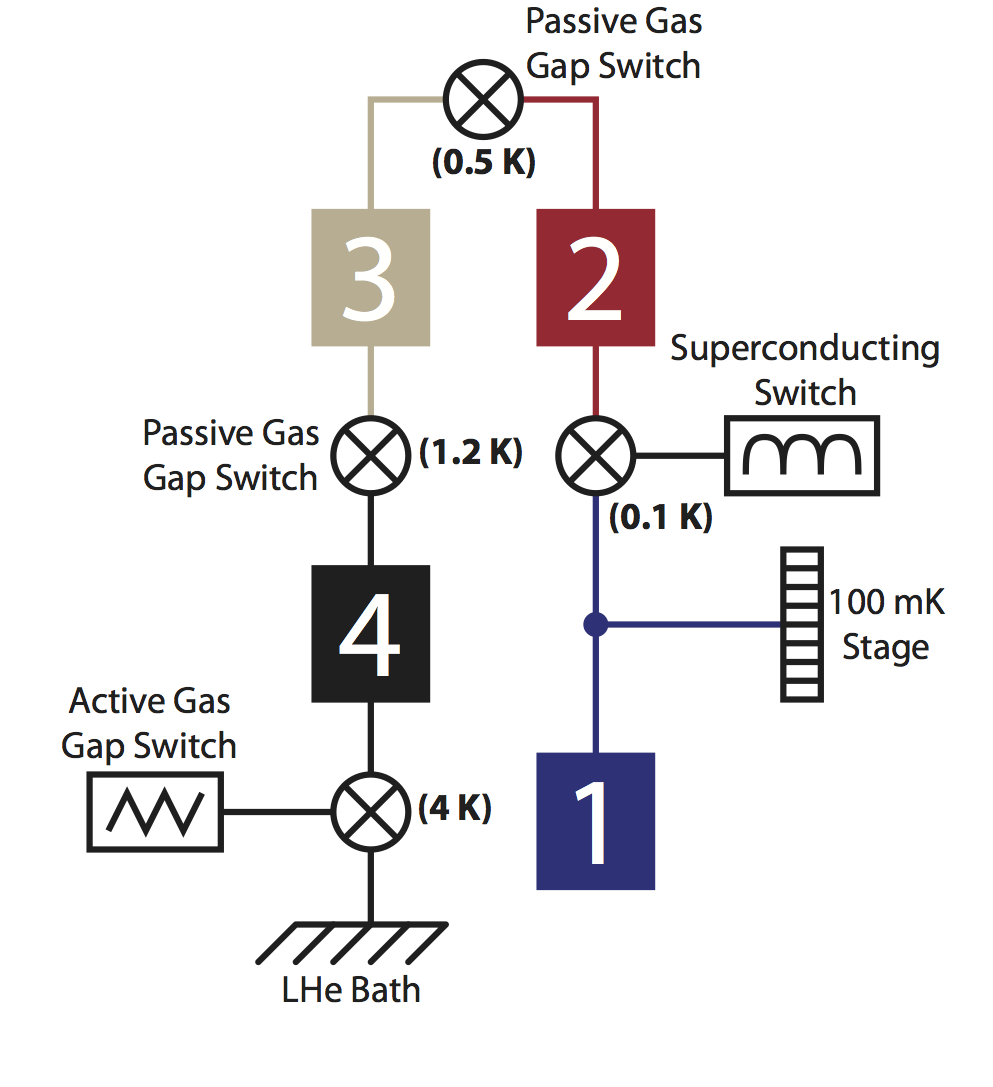}
   \includegraphics[height=6cm]{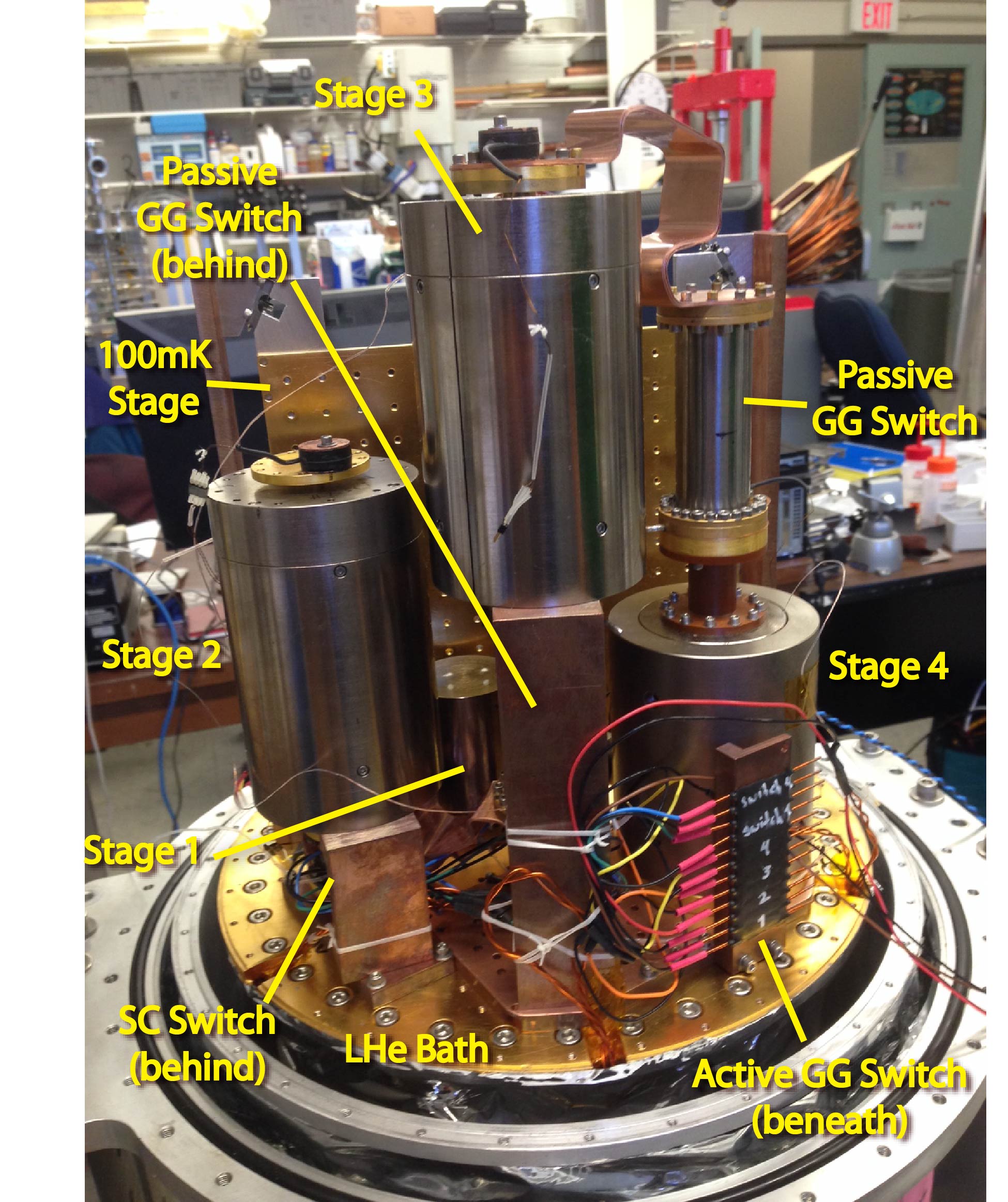}
   \includegraphics[height=6cm]{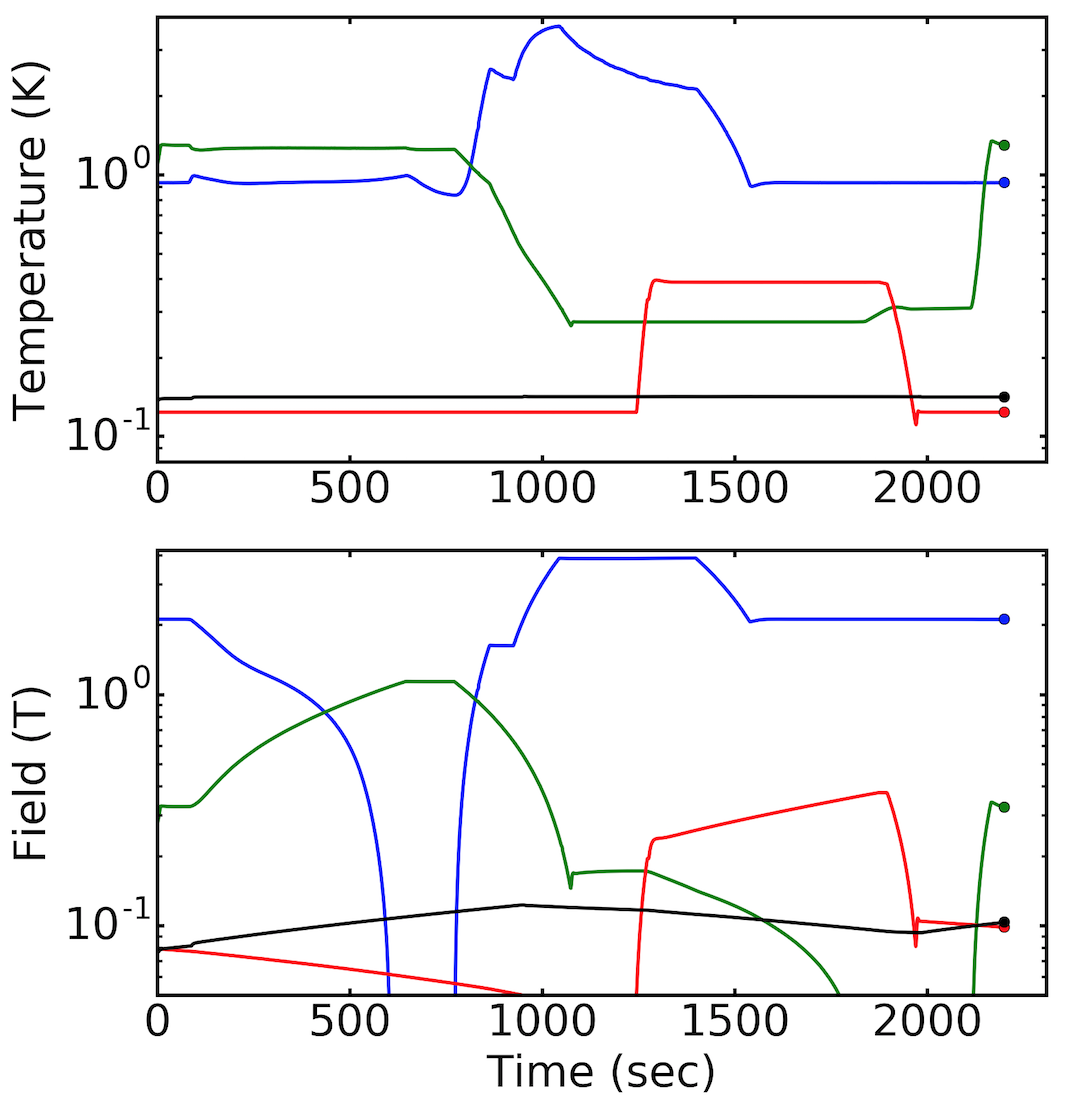}
   \end{center}
   \caption[instrument] 
   { \label{fig:ADR} Left: The 4-stage Continuous Adiabatic Demagnetization Refrigerator (CADR) provides continuous cooling power to the 100\,mK stage. Center: PIPER CADR. Right: Temperature and magnetic field versus time for the four stages during continuous operation (black: Stage 1, red: Stage 2, green: Stage 3, blue: Stage 4).}
   \end{figure}

\subsection{CRYOGENICS}\label{sec:cryo}
\,The detectors are cooled to 100\,mK using a Continuous Adiabatic Demagnetization Refrigerator (CADR) \cite{Shirron2004}. The CADR consists of four independent stages with adjacent stages connected by gas gap (GG) and superconducting (SC) heat switches (Figure \ref{fig:ADR}). Stage 1 is coupled directly to the 100\,mK stage. Stage 4 is coupled to the LHe bath through an active gas gap switch. Figure \ref{fig:ADR} shows the temperature and magnetic profiles of the four stages during operation. Stage 1 is kept at a constant temperature of 100\,mK. The upper stages demagnetize in series, allowing each of them to absorb heat from the next lower stage in a cascade up to the LHe bath. On the ground, Stage 4 cools the system from 4.2\,K to 1.5\,K. At float altitude, the LHe bath is at 1.5\,K, so Stage 4 is not used. By decoupling and cycling the various stages, the CADR can provide $20$\,$\mu$W of continuous cooling power at 100\,mK. 

\subsection{SCAN STRATEGY}
PIPER will map the sky by rotating in azimuth with the telescope mounted at a fixed elevation of $50^{\circ}$. Motion in azimuth is controlled using a Control Moment Gyro (CMG). The system consists of an identical pair of wheels spinning in opposite directions, with the spin axes perpendicular to the azimuth rotation axis. The wheels are spun up to a constant rate, then tilted about an axis perpendicular to the axis of the gondola. This transfers angular momentum to the payload, causing it to spin in azimuth. A pivot motor located at the top of the payload dumps momentum to the balloon flight train, preventing saturation which accrues if the spin axis of the wheels becomes parallel to payload spin axis.
\begin{figure}
\begin{center}
\includegraphics[height=5cm]{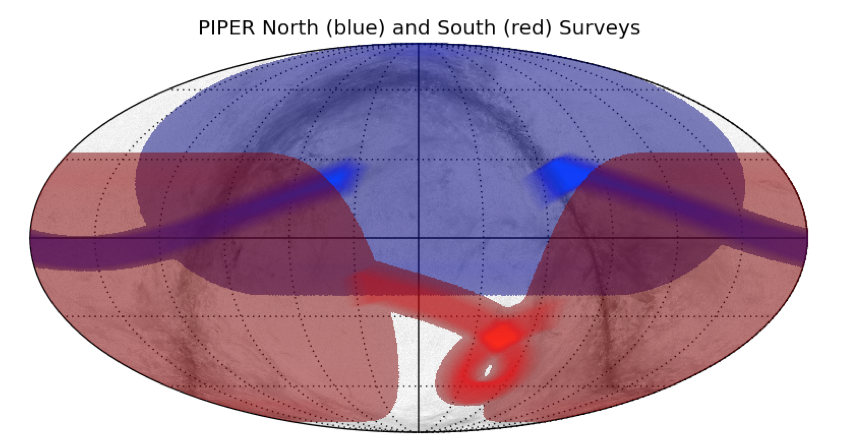}
\includegraphics[height=5cm]{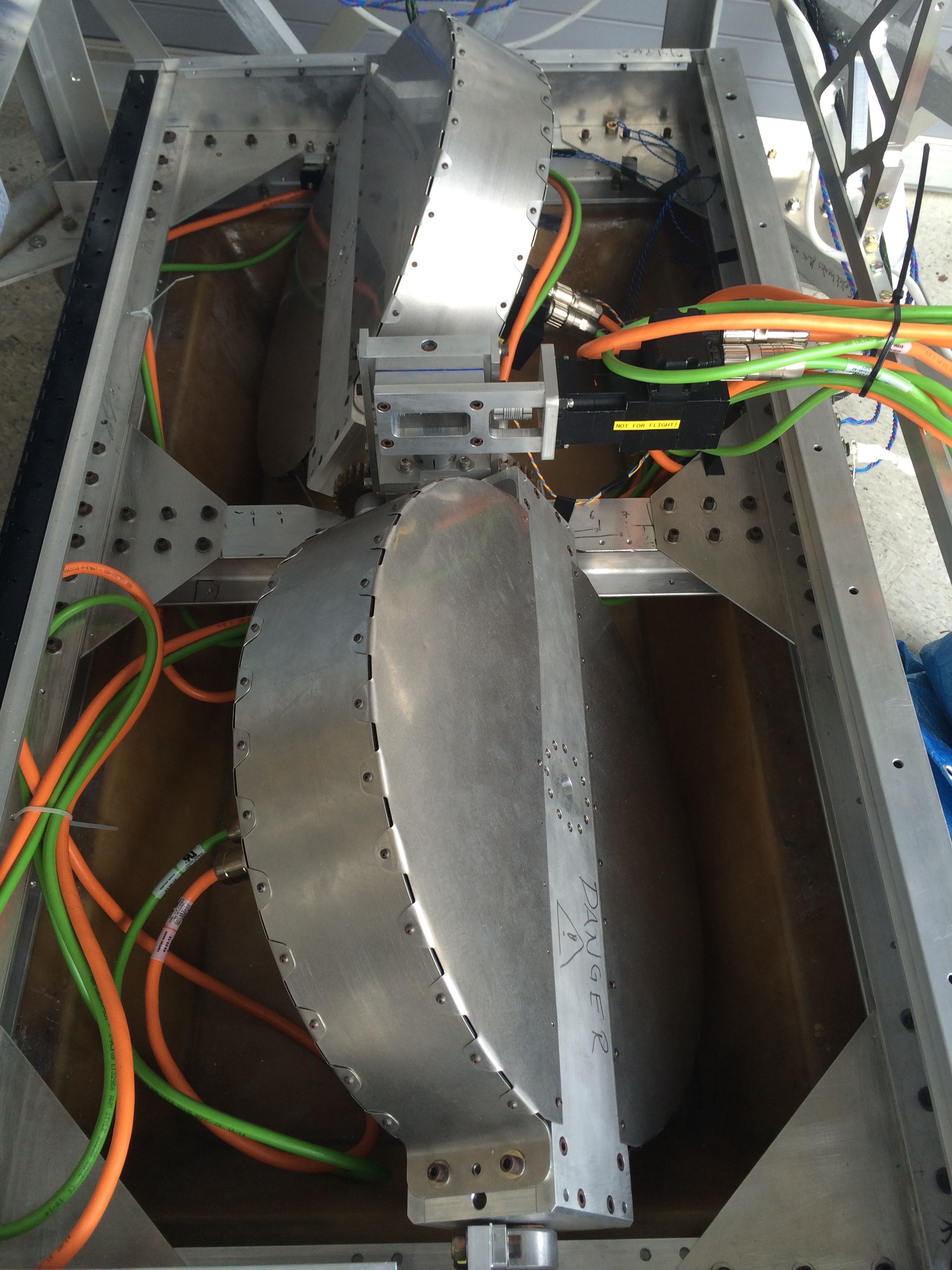}
\end{center}
\caption{Left: PIPER will map 85\% of the sky by flying from the Northern (blue) and Southern (red) hemisphere. During the night, PIPER spins in azimuth, covering the large oval regions. During the day, PIPER scans back and forth in the anti-sun direction, covering the regions shown by dark narrow strips. Right: PIPER scans in azimuth by tilting a pair of Control Moment Gyros perpendicular to their spin axes.}
\label{fig:survey}
\end{figure}

Figure \ref{fig:survey} shows the region of the sky that PIPER will observe by flying from Palestine, Texas and Alice Springs, Australia. During the night, PIPER will spin continuously at $0.5^{\circ}/s$, mapping more than 50\% of the sky in a single overnight flight. During the day it will scan a limited region of the sky, using sun sensors to avoid pointing at the Sun. The azimuthal scan speed is chosen so that the motion of the telescope beam across the sky is slower than the rate at which the VPM modulates the sky signal. In-flight attitude is determined by a combination of MEMS gyros and a magnetometer. Post-flight pointing reconstruction is provided by a boresight star tracker, which takes images of the sky every $\sim3$ seconds.

\section{CONCLUSION}
PIPER will map the polarization of the CMB in four frequency bands (200, 270, 350, and 600\,GHz) over 85\% of the sky over a series of conventional balloon flights from the Northern and Southern Hemisphere. This will allow the shape of the B-mode power spectrum to be measured at angular scales up to $90^{\circ}$, constraining the characteristic reionization bump. After eight flights PIPER will constrain the tensor-to-scalar ratio to $r<0.007$. PIPER will also provide a high signal-to-noise map of the polarized emission from galactic dust, producing a high fidelity template that can be used by CMB experiments to remove the foreground signal from their measurements.

%\acknowledgments % equivalent to \section*{ACKNOWLEDGMENTS}       
 
%This unnumbered section is used to identify those who have aided the authors in understanding or accomplishing the work presented and to acknowledge sources of funding.  

% References
\bibliography{report} % bibliography data in report.bib
\bibliographystyle{spiebib} % makes bibtex use spiebib.bst

\end{document}